\documentclass[final,english]{bullsrsl}[2022/06/15]

\usepackage[latin1]{inputenc}
\usepackage[T1]{fontenc}

\usepackage{natbib} 
\usepackage{graphicx}

\begin{document}
\title{Synchrotron radio emission as a proxy to identify long period massive binaries}

\author[corresponding]{Micha\"el}{De Becker}
\author[]{Bharti}{Arora}
\affiliation{Space sciences, Technologies, and Astrophysics Research (STAR) institute, University of Li\`ege, Belgium}
\correspondance{Michael.DeBecker@uliege.be}
\date{30th April 2023}
\maketitle

\begin{abstract}
The multiplicity of massive stars is known to be significantly high. Even though the majority of massive stars are located in binary systems, the census of binaries is biased toward shorter periods as longer period systems are more difficult to identify. Alternatively, the search for binary systems with longer periods may proceed differently. As massive binary systems are typically colliding-wind systems, hints for processes occurring in the colliding-wind region could be used as a valuable proxy to identify likely binary systems, and then organize dedicated spectroscopic or interferometric campaigns on a short list of pre-selected targets. In this context, any hint for synchrotron radio emission is seen as a promising indicator of long period binaries, as short period systems undergo severe free-free absorption of the synchrotron emission by the stellar wind material. Usual techniques to identify synchrotron radio emitters constitute thus valid tools to explore that poorly investigated part of the massive binary parameter space. In addition, the identification of a synchrotron emission component in a short period binary can be used as an indicator of the presence of a third companion on a still unrevealed wider orbit in a triple system. 
\end{abstract}


\section{The multiplicity of massive stars}\label{multi}
\begin{figure}
\centering
\includegraphics[width=12cm]{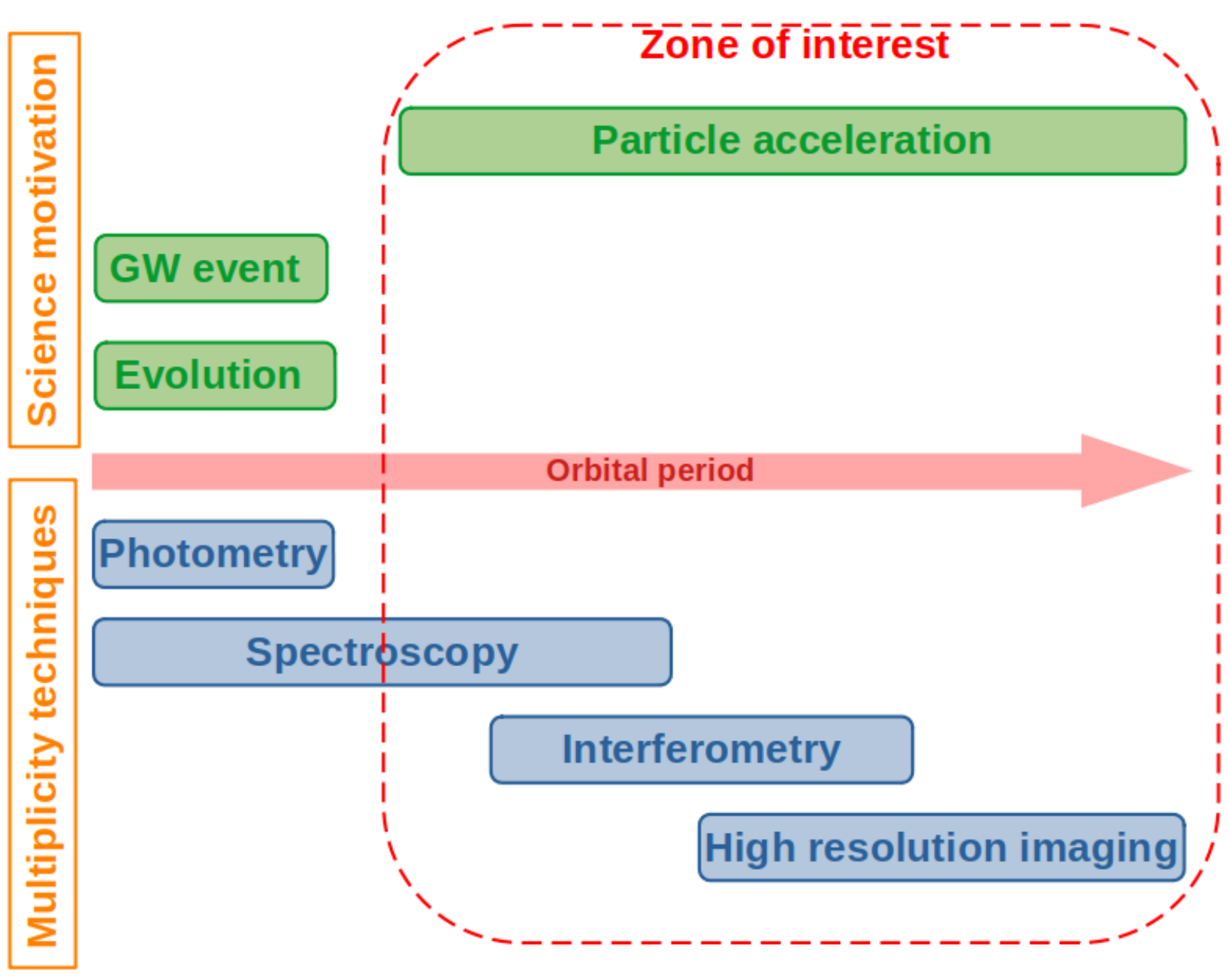}
\begin{minipage}{12cm}
\centering
\caption{Diagram illustrating the main science drivers motivating the study of massive binaries, along with the main techniques used to investigate their multiplicity as a function of the orbital period.\label{Fig1}}
\end{minipage}
\end{figure}

A large fraction of massive stars are part of binary, or higher multiplicity systems \citep{Sana2012,OffnerMultiplicity}. However, the census of binary systems is dependent on the specificities of the direct techniques used to investigate their multiplicity (see blue boxes in Fig.\,\ref{Fig1}),
\begin{enumerate}
\item[-] Shorter time scales are easier to monitor as the duration of the time series is shorter, and observation campaigns are easier to organize.
\item[-] Photometry and spectroscopy are well-known techniques easy to implement, and especially efficient at unveiling and characterizing short period systems (typically a few days, even though spectroscopy can be efficient up to a few year). However, these techniques are valid provided the inclination of the system is appropriate. The measurement of eclipses through photometric time series requires indeed the line-of-sight to be very close to the orbital plane. An orbital plane seen under a not too low inclination angle is also a requirement for the identification and study of spectroscopic binaries. The Doppler shift of individual stellar spectral lines results from the periodic motion along the line-of-sight of the components of the binary system (either approaching or receding from the observer). A pole-on viewed orbit is thus not able to display any radial velocity shifts along the orbital motion. In addition, long period systems display typically lower amplitude radial velocity curves, and their high eccentricity leads radial velocities not to vary in a spectacular way over a large fraction of the orbit. 
\item[-] Long baseline interferometry is a complementary technique very useful to investigate periods of the order of one year, up to a few tens of years \citep{LeBouquin2017,Lanthermann2023}. However, only a couple of facilities exist (e.g., the Very Large Telescope Interferometer at the European Southern Observatory, Chile, or the Center for High Angular Resolution Astronomy array on Mount Wilson, California), and this is not enough to organize significant surveys.
\item[-] Non-interferometric high resolution imaging is relevant for much longer periods, but such long time-scales are not easy to monitor \citep{Sana2014}.
\end{enumerate}
Given the constraints listed above, long period systems are still poorly investigated, leading to a significant bias in the exploration of the massive binary parameter space. 

\section{Sciences drivers as a function of the orbital period}
Depending on the orbital period of massive binaries, various science drivers can be identified (see green boxes in Fig.\,\ref{Fig1}). Shorter period systems are privileged targets for modern stellar astrophysics for various reasons, including in particular these ones: 
\begin{enumerate}
\item[1.] They constitute potential progenitors for gravitational wave (GW) events triggered by black hole mergers or neutron star mergers \citep{GWcata2019}. 
\item[2.] Mass exchange between components of short period binary systems is a key aspect of massive star evolution \citep{Sana2012}.
\end{enumerate}
On the other hand, long period systems are relevant laboratories to investigate shock physics. The stellar winds of both stars collide at high speeds, typically their terminal velocity ($V_\infty$ = 2000 -- 3000 km\,s$^{-1}$), leading to high Mach number adiabatic shocks. A specific topic of interest in these systems is particle acceleration through the Diffusive Shock Acceleration process \citep{Drury1983}. This process is known to accelerate charged particles, including electrons, up to relativistic velocities. This population of relativistic electrons is likely to participate in non-thermal emission processes. In particular, in the presence of the magnetic field of stars in the system, one can expect the synchrotron emission process to be active in the radio domain \citep[e.g.][]{DeBecker2007}.  

\section{Synchrotron radiation as an tracer of long period massive binaries}
Most particle accelerators among massive binaries, referred to as Particle-Accelerating Colliding-Wind Binaries (PACWBs, \citealt{pacwbcata}\footnote{\url{https://www.astro.uliege.be/~debecker/pacwb/}}) are identified through the detection of synchrotron radio emission from the colliding-wind region \citep[e.g.][]{Abbott1984,Benaglia2006,Dougherty2005,Blomme2007,Benaglia2015,Benaglia2020,Marcote2021}. Let's stress that single massive stars do not offer the opportunity to accelerate particles, and thus to produce synchrotron radiation, except in the scarce situation of a powerful wind interacting with a sufficiently dense interstellar medium (the only known case to date is of WR102, \citealt{Prajapati2019}). {\it Particle acceleration and synchrotron emission by massive stars is strongly correlated to their multiplicity}, as confirmed by the current census of PACWBs. However, the synchrotron emission region can be significantly embedded in the stellar wind plasma. Depending on the wind parameters, on the orbital parameters and on the orbital phase, the synchrotron emission can be significantly attenuated by free-free absorption (FFA). This process is especially active in short period binaries where the wind collision occurs in a denser plasma. PACWBs are thus more likely revealed in longer period systems (typically, at least several weeks, but generally more than a few months). 

As a result, one can claim that {\it not all massive binaries are able to reveal some synchrotron radio emission, but all synchrotron-emitting massive stars are part of a binary or higher multiplicity system}. This fact allows to consider that the firm identification of a synchrotron component in the radio emission from massive stars is telling us about their multiplicity.

\section{General strategy}
Even though large scale surveys of stellar populations can turn out to be quite useful to investigate the multiplicity of massive stars, it will most of the time require to gather long-term time series. This is quite demanding in term of telescope time and, as stated in Sect.\,\ref{multi}, observational biases are strongly affecting most techniques implemented for such an endeavour. Alternatively, direct methods such as spectroscopic and high angular resolution imaging techniques can be used following a more efficient approach, focusing on target lists made of stars whose multiplicity has been revealed by indirect methods, such as identifying them as synchrotron emitters. 

The radio measurement of many massive stars not known to be long period binaries is thus expected to reveal synchrotron radiation from some of them, according to well-established interpretation guidelines \citep{pacwbcata,DeBecker2017}: variability studies, high-angular resolution measurements, spectral index measurements based on at last two spectral bands, search for high brightness temperature radio emission. If synchrotron radiation is identified, it points to a binary system with a period long enough to warrant that FFA is not too severe. From this point, direct techniques can be applied to pre-selected targets, aiming for the detection of the companion, the identification of its spectral type and the determination of the orbital parameters of the system.

This approach deserves also to be considered for known short period binaries, as it is likely to reveal triple systems with a third companion on a much wider orbit. This approach has already been quite useful to confirm the triple system status of a few massive star systems whose short orbit was too small (typically a few days) to explain the detection of synchrotron radio emission. One may mention the case of HD\,167971 whose wide orbit was revealed by long baseline interferometry \citep{DeBecker2012}, or that of HD\,150136 investigated by using spectroscopic and interferometric techniques \citep{Sana2013}. More recently, the careful analysis of the radio emission from the 112-d binary WR133 led to the conclusion that it could be a triple system, with a still undetected third object on an orbit of potentially a few years \citep{DeBecker2019}.

\section{Summary and concluding remarks}

The key points addressed in this paper are the following:
\begin{enumerate}
    \item[$\bullet$] Given the difficulty to reveal long period binaries with usual direct techniques, the census of long period systems is underestimated. As a result, complementary indirect approaches to reveal long period systems are welcome.
    \item[$\bullet$] An important feature of massive binaries is their capability to allow their stellar winds to collide. These wind collisions are appropriate to drive shock physics, including particle acceleration.
    \item[$\bullet$] PACWBs are mainly identified through their synchrotron radio emission. One can thus consider this non-thermal radio emission signature as an indicator of binarity.
    \item[$\bullet$] Synchrotron radio emission is easily attenuated by free-free absorption that is especially strong in short period systems. The synchrotron signature is thus expected to be revealed provided the orbital period is not too short, as the colliding-wind region is not too deeply embedded in the dense part of the stellar winds. As a result, synchrotron radio emission and particle acceleration can be considered as good indirect indicators of binary systems with rather long periods.
    \item[$\bullet$] Once a candidate binary system is identified thanks to its synchrotron radio signature, a more focused use of direct multiplicity study techniques (such as spectroscopy or long baseline interferometry) can be envisaged to characterize the newly discovered systems. In particular, the identification of the companion's nature and the determination of orbital elements cannot be achieved by the sole use of radio measurements.
\end{enumerate}

It is important to stress that such an approach also turns out to be efficient at revealing the existence of a third companion in already known binary systems. Typically, a massive star system known as a very short period binary and displaying significant synchrotron radio emission could actually be a triple system, with the non-thermal radio emission arising from the collision region produced by the interaction of the stellar wind of a third star with that of the previously known close binary. Several members of the catalogue of PACWBs were confirmed as binaries or higher multiplicity systems thanks to specific multiplicity studies motivated by their particle accelerator status. The use of their synchrotron radio emission is thus a relevant approach to investigate the multiplicity of massive stars in general.

Beside the interest of this approach to unveil new particle accelerators to be discussed in the context of the role played by PACWBs in the acceleration of galactic cosmic rays \citep[see][in this volume]{DeBecker2023}, a better census of the multiplicity of massive stars is of high importance for the question of their formation and dynamic evolution in stellar clusters.

\begin{acknowledgments}
The authors would like to thank the organizers for the opportunity to present this poster, along with the Belgo-Indian Network for Astronomy and Astrophysics (BINA) for financial support. Bharti Arora acknowledges financial support from Wallonia-Brussels International (WBI). This research has made use of NASA's Astrophysics Data System Bibliographic Services. This work is supported by the Belgo-Indian Network for Astronomy and astrophysics (BINA), approved by the International Division, Department of Science and Technology (DST, Govt. of India; DST/INT/BELG/P-09/2017) and the Belgian Federal Science Policy Office (BELSPO, Govt. of Belgium; BL/33/IN12).
\end{acknowledgments}

\begin{furtherinformation}

\begin{orcids}
\orcid{0000-0002-1303-6534}{M.}{De Becker}

\end{orcids}

\begin{authorcontributions}
Both co-authors actively contributed to this work. 
\end{authorcontributions}

\begin{conflictsofinterest}
The authors declare no conflict of interest.
\end{conflictsofinterest}

\end{furtherinformation}

\bibliographystyle{bullsrsl-en}

\bibliography{extra}

\end{document}